\documentclass[reprint, amsmath, amssymb, groupaddress, aps, prb]{revtex4-1}
\usepackage{graphicx}
\usepackage{amsmath}
\begin{document}
\title{Encoding Quantum Mechanics into Pentagonal Geometries for Discovering Novel Two-Dimensional Materials}
\author{Lei Liu}
\author{Immanuella Kankam}
\author{Houlong L. Zhuang}
\email{zhuanghl@asu.edu}
\affiliation{School for Engineering of Matter Transport and Energy, Arizona State University, Tempe, AZ 85287, USA}
\date{\today}
\begin{abstract}
Imagine that there is a gapless plane tessellated by irregular, convex pentagons with their side lengths at the sub-nanoscale, and tiny balls are placed at the vertices of each pentagon. If there are no interactions among these balls, one would expect that they stand still at the vertices of the pentagons. But what if these balls symbolize atoms? The electrons and nuclei of the atoms will start to interact following the laws of quantum mechanics. As a result, the atoms will relocate leading to a system of energy minimum. While performing an experiment of manipulating individual atoms and observing the resulting configurations may be challenging, we apply {\it ab initio} density functional theory (DFT) calculations to carry out a virtual experiment, where we assign carbon atoms at the vertices of 14 types of convex pentagons and then optimize the positions of these atoms. We obtain a variety of unexpected structures; almost all of them except types 2 and 4 structures--these two structures are equivalent--differ from the initial pentagonal geometries. We compute the band structure of each resulting geometry representing the crystal structure of a two-dimensional (2D) material. We find diverse electrical properties among these 2D materials: metallic, semiconducting, and semi-metallic with a Dirac cone. Our work shows that encoding quantum mechanics into the geometries of irregular, convex pentagons and with the help of DFT calculations open up a novel route for accelerating discovery of new 2D materials.
\end{abstract}
\maketitle
\section{Introduction}
\begin{figure*}
  \includegraphics[scale=0.22]{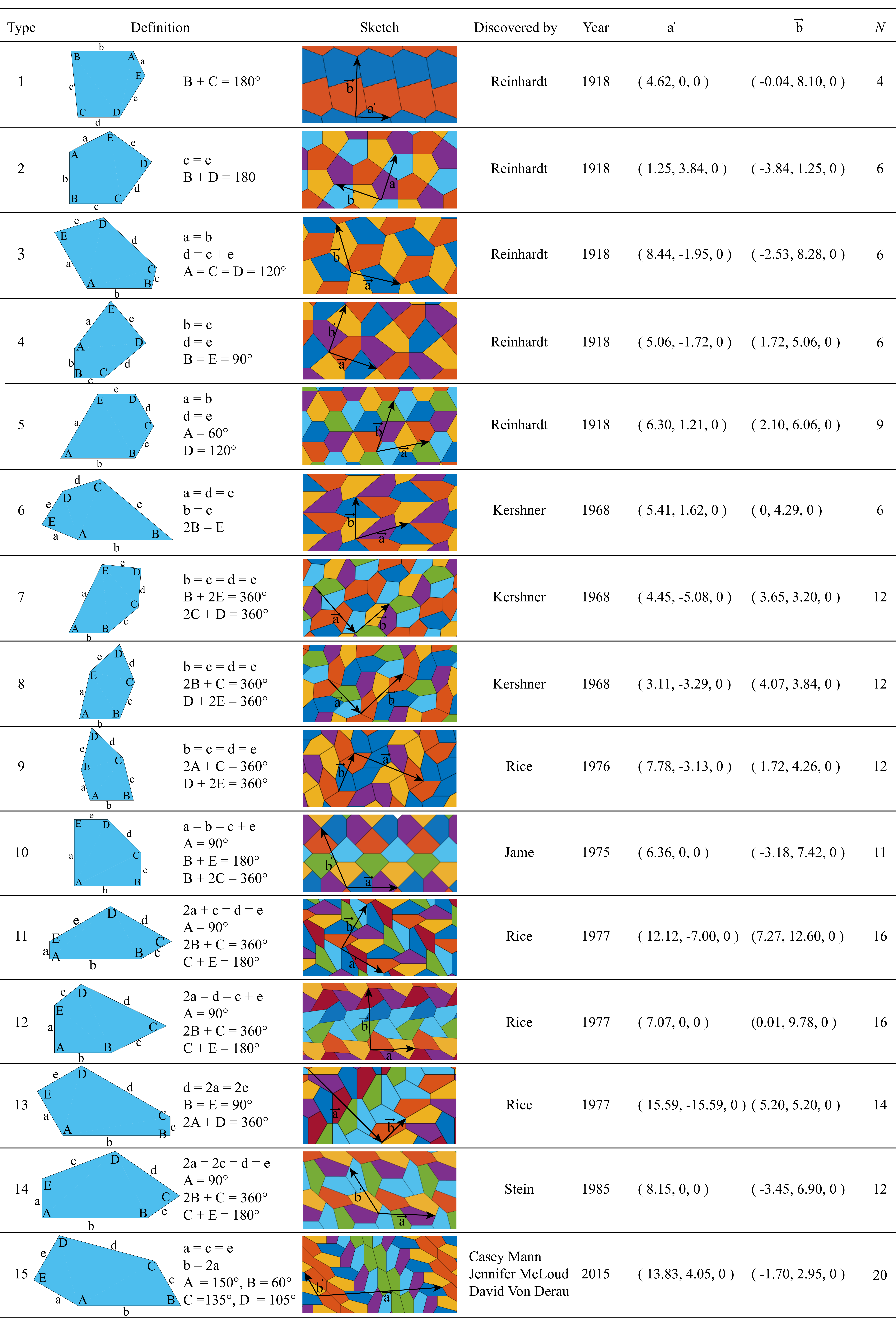}
  \caption{Sketches of 15 types of tessellating convex pentagons and their color-filled patterns in a plane. The definition and sketches of the pentagons are adopted from Refs.~\onlinecite{kershner1968paving,wells1991penguin}. The discoverers and years of the pentagons are also shown. The translational vectors of the pattern are not unique. We choose these translational vectors, whose components have units of \AA, to be the in-plane lattice vectors $\protect\overrightarrow{a}$ and $\protect\overrightarrow{b}$ of a two-dimensional (2D) material. The number of atoms in the unit cell of a 2D material based on one of the 15 types of pentagons is denoted as $N$.}
  \label{fig:sketch}
\end{figure*}

Pentagonal geometries are attractive to professional and amateur mathematicians, for whom one long-standing question is how many types of irregular, convex pentagons can tessellate a plane leaving no gaps or overlaps. It is well known that a regular pentagonal tiling on the Euclidean plane is impossible because its internal angle is not a divisor of 360$^\circ$. But the number of families of irregular, convex pentagons that tile a plane have grown to 15 since the discovery of the first type by Reinhardt in 1918.\cite{wells1991penguin} Figure~\ref{fig:sketch} illustrates 15 types of pentagons that have been discovered so far. This list is recently claimed to be complete according to the proof by Rao.\cite{rao2017exhaustive}

The 15 types of pentagons are distinguished from each other by the constraints of the five angles and the five side lengths. For example, the sum of angles B and E in type 10 pentagon is 180$^\circ$, whereas there is no such a geometrical relation in type 3 pentagon. Each of these 15 monohedral tiling types possesses more than one degree of freedom, except types 14 and 15 whose interior angles are all fixed. Apart from different constraints, one can distinct these 15 types by their primitive units and symmetry groups of tiling topologies. For example, the primitive cell of type 13 tiling plane contains eight pentagons while six pentagons make up the primitive cell of type 14 tiling plane.

Tessellating a plane with irregular, convex pentagons is like playing a jigsaw puzzle. However, encoding the puzzle with laws of quantum mechanics by placing atoms at the vertices of each piece of the puzzle will likely alters the shape of the piece owing to the electron-electron, electron-ion, and ion-ion interactions. We will show that {\it Ab initio} density functional theory (DFT) calculations can be used to play this imaginary puzzle.

{\it Ab initio} DFT simulations are an ideal tool to relax the atomic positions of a system until it reaches a configuration of local energy minimum of the corresponding high-dimensional energy surface. Essentially, DFT is an eigenvalue solver giving information of total energy that can be used to derive a number of properties of the simulated system. What is needed for DFT calculations is often as simple as the lattice parameters and atomic coordinates of a crystal or molecule. 

To connect pentagonal geometries with DFT simulations, we analogized an infinite plane of pentagons to the periodic crystal structure of a two-dimensional (2D) material in our recent theoretical study.\cite{liu2018} We applied DFT calculations to examine the possibility of forming a nanosheet filled up by type 15 pentagons using eight different elements. Unfortunately, no atoms of any of the eight elements were found to be able to remain at the vertices of a plane of type 15 pentagons. On the contrary, we obtained a variety of non-pentagonal geometries depending on the elements.  

In this work, we extend our DFT study to other 14 types of pentagons. Instead of studying different species, we focus on a single element, carbon, which can form a number of polymorphs with flexible bonding environments. For example, in bulk form, two of the most familiar polymorphs of carbon are graphite and diamond. In addition, Zhang {\it et al.} recently have predicted a new 3D form with interlocking hexagons.\cite{zhang2013stable} In the regime of nanomaterials, carbon atoms can form 0D fullerenes C$_{60}$ and C$_{70}$,\cite{kroto1985c60, buseck1992fullerenes} 1D carbon nanotube,\cite{iijima1991helical} and 2D graphene.\cite{novoselov2004electric} We therefore expect carbon atoms may be able to arrange themselves at the vertices of any of the 14 types of pentagons.  

\section{Methods}
\begin{figure}
  \includegraphics[scale=0.3]{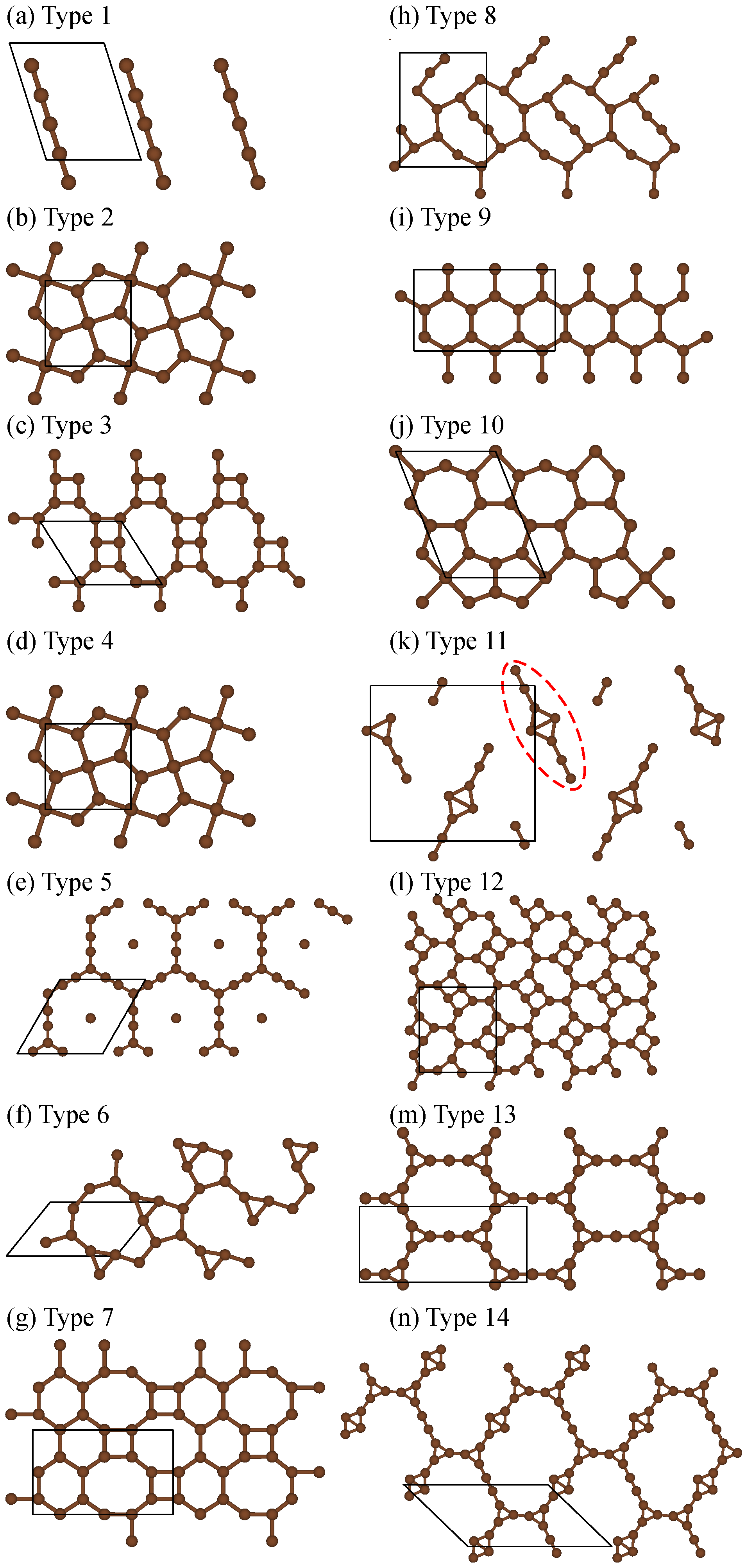}
  \caption{Optimized structures of carbon nanosheets with their initial structural models created based upon the 14 types of convex pentagons. Unit cells of the carbon nanosheets are enclosed by the solid lines. An isolated cluster in the structure model of (k) is highlighted with a red dashed ellipse.}
  \label{fig:structure}
\end{figure}

All the density functional theory (DFT) calculations such as geometry optimizations and electronic structure are performed using the Vienna {\it Ab-initio} Simulation Package (VASP, version 5.4.4).\cite{Kresse96p11169} The Perdew-Burke-Ernzerhof (PBE) functional within the generalized gradient approximation is applied to treat the exchange-correlation interactions.\cite{Perdew96p3865} The electron-ion interactions are described by the PBE version of the potential dataset generated based on the projector-augmented wave (PAW) method.\cite{Bloechl94p17953,Kresse99p1758} We adopt surface supercell models to simulate each carbon nanosheet with the C atoms placed at the vertices of each type of pentagon. Note that several types (types 1, 2, 3, 4, 5, 6, and 8) of pentagons contain more than one tiling topologies. Here we use one of the most basic representations for these types of pentagons. Figure~\ref{fig:sketch} shows the lattice vectors and the number of atoms in each unit cell. These lattice vectors are chosen to ensure that the initial nearest distance between neighboring C atoms is at least 1.4~\AA. To satisfy the periodic boundary conditions (PBC) in all the three directions, we add a vacuum layer with the thickness of 18.0~\AA~ between the neighboring carbon nanosheets. This thickness is sufficiently large that image interactions between surface supercells are negligible. Benefited from the PBC, plane waves with the kinetic cutoff energy below 600~eV are included to approximate the total electron wave function. We employ a $\Gamma$-centered $9~\times~9~\times~1$ $k$-point grid for the integration in the reciprocal space.\cite{PhysRevB.13.5188} Furthermore, we apply the conjugate-gradient algorithm\cite{Shewchuk} as implemented in VASP to optimize the structures of carbon nanosheets. The in-plane lattice parameters and atomic positions are completely optimized until the residual inter-atomic forces are smaller than 0.01 eV/\AA.
\section{Results and Discussion}
\begin{table}[b]
  \caption{Properties of 14 types of carbon nanosheets: the optimized lattice constants ($\|a\|$, $\|b\|$, and $\alpha$), the shortest C-C bond length $d$, the energy difference $\Delta E$, and the bandgap $E_\mathrm{g}$ calculated with the PBE functional. $\Delta E$ is calculated with reference to that of type 9 carbon nanosheet, {\it i.e.,} graphene. Metallic carbon nanosheets are labelled as M.}
  \begin{ruledtabular}
    \begin{center}
      \begin{tabular}{ccccccc}
            Type & $\|a\|$ (\AA) &$\|b\|$ (\AA)  &$\alpha~(^\circ)$  &$d$ (\AA) &$\Delta E$ (eV/atom) & $E_\mathrm{g}$(eV) \\
            \hline
1&3.96 &5.13 &107.53 &1.26  &0.99 &0.39\\
2&3.92 &3.92 &90.00 &1.37 &1.76 &M \\
3&4.97 &4.61 &122.65 &1.36 &0.94 &0.14 \\
4&3.92 &3.92 &90.00 &1.37 &1.75 &M \\
5&6.95 &6.95 &60.00 &1.23 &1.82 &M \\
6&6.15 &3.89 &50.81 &1.39 &0.91 &M \\
7&7.52 &4.53 &90.00 &1.34 &0.44 &M \\
8&5.14 &6.76 &90.00 &1.20  &1.55 &0.46\\
9&7.40 &4.27 &90.00 &1.42  &0 &M\\
10&4.84 &6.59 &111.55 &1.38 &0.68 &M \\
11&10.17 &9.63 &90.00 &1.28 &2.02 &0.89 \\
12&6.44 &7.11 &90.00 &1.40 &0.85 &0.25 \\
13&11.51 &5.23 &90.00 &1.31 &1.19 &M \\
14&12.23 &7.56 &136.00 &1.25  &1.23 &M\\
      \end{tabular}
    \end{center}
  \end{ruledtabular}
  \label{summary}
\end{table}

Figure~\ref{fig:structure} shows the structures of the 14 types of carbon nanosheets optimized from the 14 types of pentagons as the input geometries for DFT structural relaxations. As can be seen, type 1 carbon nanosheet consists of linear carbon chains in the direction of lattice vector $\overrightarrow{b}$. The distance (3.96~\AA) between the chains may be large enough to separate interactions between neighboring chains. Type 2 carbon nanosheet, interestingly, is able to retain the shape of type 2 pentagons. Namely, the DFT structural optimizations merely leads to a linear scaling of lattice parameters and to a reduced nearest C-C distance of 1.37\AA. Type 3 carbon nanosheet exhibits pairs of four-membered rings. Type 4 carbon nanosheet is equivalent of type 2 carbon nanosheets, although their initial pentagons belong to two different types. Type 5 carbon nanosheet shows a pattern of large hexagons, where almost all of the C atoms locate at the vertices or edges. The exception is that one isolated C atom is found at the centers of the hexagons. Type 6 carbon nanosheet has pairs of three and five-membered rings connected by curved chains of carbon atoms. The structure of type 7 carbon nanosheet appears regular, containing a pattern of square, hexagon, and octagons with C atoms occupying their vertices. By contrast, C atoms in type 8 carbon nanosheet settle at the vertices distorted octagons. It is straightforward to identify type 9 carbon nanosheet with six-membered rings as graphene, and the unit cell the nanosheet is just a different representation of graphene unit cell. Note that type 15 carbon nanosheet in our previous study is also found to be equivalent to graphene.\cite{liu2018} Type 10 carbon nanosheet consists of pairs of five- and seven-membered rings. These pairs five-seven membered rings are not uncommon, as they are typically found at graphene grain boundaries.\cite{huang2011grains} Atomic models based on these pairs have been commonly used to study a variety of properties such as thermal transport properties of polycrystalline graphene.\cite{bagri2011thermal} Type 11 carbon nanosheet shows a pattern of isolated clusters. Each of the cluster has eight C atoms aligned in a nearly linear chain with two of them in the middle protruded out of the chain. The structure of type 12 carbon nanosheet is similar to that of type 3 carbon nanosheet comprising pairs of four-membered rings. But these rings in type 12 carbon nanosheet are arranged in a pattern of rotation. C atoms in type 13 carbon nanosheet form three-membered rings surrounding a nearly circular chain of C atoms. In type 14 carbon nanosheet, carbon atoms also form three-membered rings as in type 13 carbon nanosheet. In addition, a pair of rings share the edge. All the rings are connected by a curved chain of C atoms. Taking into account equivalent types 2 and 4, and types 9 and 15 carbon nanosheets, there are only 13 types of inequivalent carbon nanosheets, in contrast to 15 types of inequivalent pentagons. These 13 different structures of carbon nanosheets reflect the diversity and flexibility of C-C bonding.  

Table~\ref{summary} summarizes the in-plane lattice constants and the shortest C-C bond length for each of the 14 carbon nanosheets. The angles between the two lattice vectors of types 2, 4, 7, 8, 9, 11, 12, and 13 carbon nanosheets show that the surface slabs for simulations remain orthorhombic after geometry optimizations. Typical bond lengths of C-C single, double, and triple bonds are 1.53, 1.34, and 1.20~\AA~as in C$_2$H$_6$, C$_2$H$_4$, and C$_2$H$_2$ molecules,\cite{wells2012structural} respectively. The C-C bond lengths of the 14 carbon nanosheets are within the range spanned by the three types of C-C bonds, suggesting mixed bonding types in the 14 carbon nanosheets. 

\begin{figure}
  \includegraphics[scale=0.35]{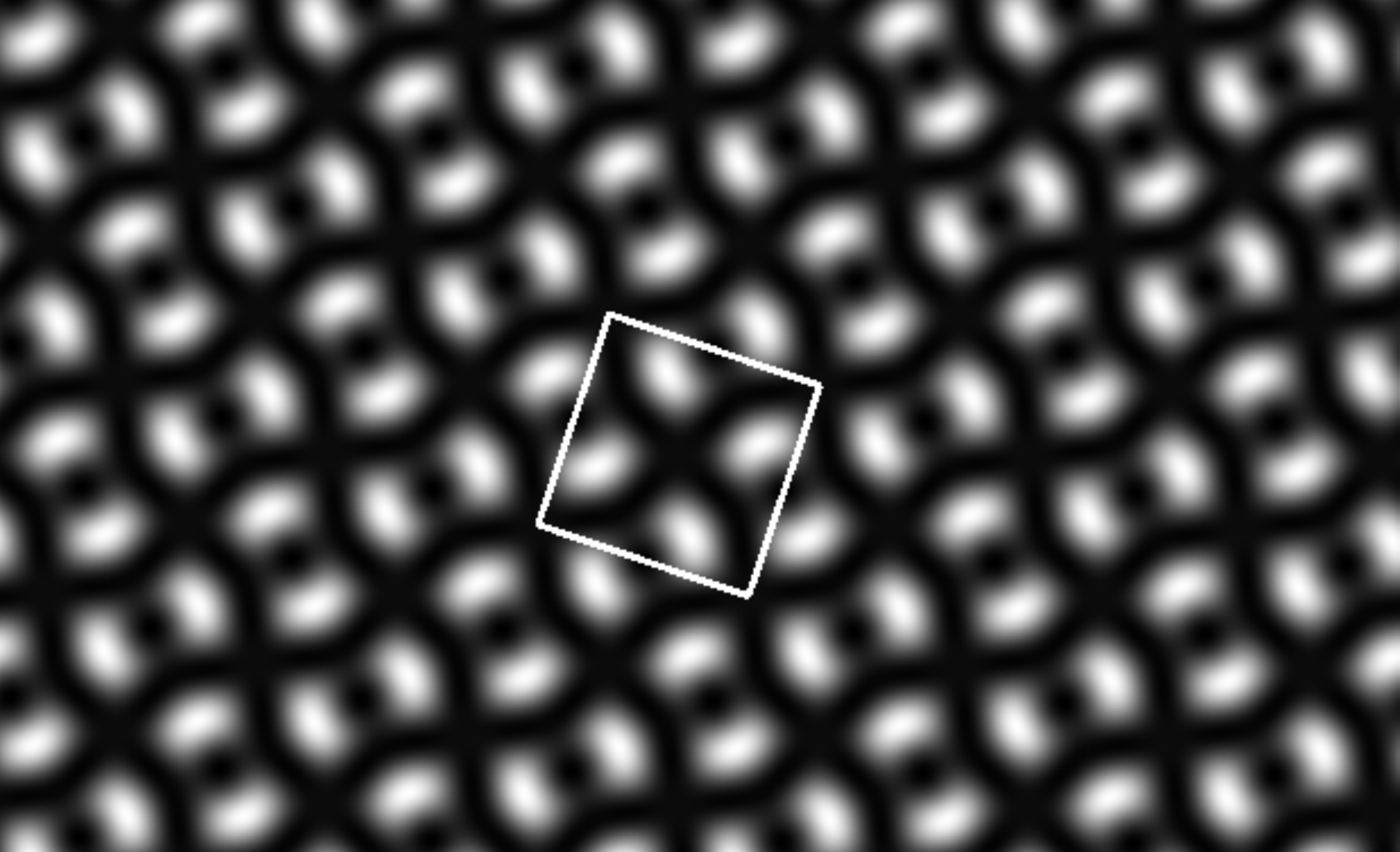}
  \caption{Simulated scanning tunneling microscope image of type 2 or 4 carbon nanosheet. The solid white lines encompass a unit cell.}
  \label{fig:stm}
\end{figure}

Because type 2 (or 4) carbon nanosheet is the only structure with C atoms that can remain at the vertices of type 2 pentagon, we perform simulations of scanning tunneling microscope (STM) on this particular nanosheet, as STM is a robust technique to investigate atomic-scale surface structures.\cite{PhysRevLett.49.57} We employ the Tersoff-Haman approach for simulating STM images.\cite{PhysRevB.31.805}  Partial charge density with an energy window between -0.2 and 0.2 eV with reference to the Fermi level is used for the calculations. Figure~\ref{fig:stm} shows the simulated STM image under the constant-height mode with the STM tip at 1.0~\AA~above the carbon nanosheet. The simulated STM image may facilitate future experimental confirmation of the unique pentagonal structure of type 2 carbon nanosheet.

To evaluate the energetics of the 14 carbon nanosheets, Table~\ref{summary} also shows their relative energies. We find that type 9 carbon nanosheet ({\it i.e.,} graphene) exhibits the lowest energy revealing that six-membered rings are the most energetically favorable bonding environment in 2D. We therefore set the energy of type 9 carbon nanosheet to zero and as a reference to calculate the relative energies of the other 13 carbon nanosheets. The relative energies are generally large, ranging from 0.44 eV/atom to 2.02 eV/atom. Interestingly, type 7 carbon nanosheet exhibits the lowest relative energy (0.44 eV/atom), possibly because of the six-membered rings embedded in the crystal structure as can seen in Fig.\ref{fig:structure}(g). Each C atom in this type of nanosheet is also surrounded by three C atoms, a configuration that resembles graphene the most. The high relative energies indicate significant energy costs are required to form the 13 carbon nanosheets. In other words, it is most unlikely to obtain isolated carbon nanosheets with the 13 structures. However, we expect that a small area of these nanosheets could exist, for example, in grain boundaries of graphene nanosheets. Because once these non-graphene nanosheets form during the growing process of graphene on a metal substrate by the chemical vapor deposition method,\cite{zhang2013review} the energy barrier of transforming the other carbon rings to six-membered carbon rings may be too large to occur. 

\begin{figure}
  \includegraphics[scale=0.27]{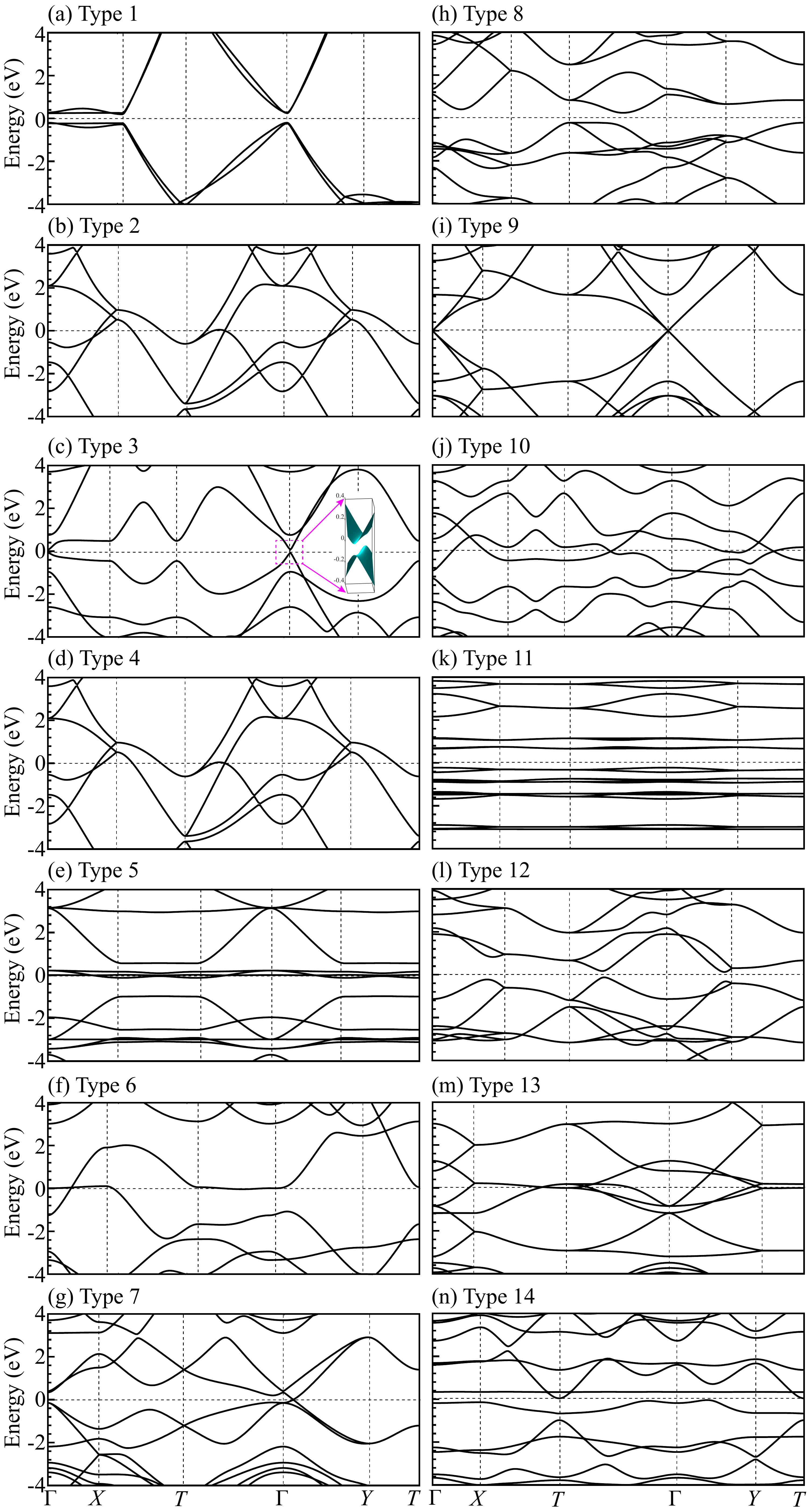}
  \caption{Predicted band structures of 14 carbon nanosheets with their crystal structures optimized from initial geometries of 14 distinct types of convex pentagons. The band structures are calculated at the DFT-PBE level. The high-symmetry $k$ points dividing the chosen path from the left to the right into five segments  have fractional coordinates of $\Gamma$(0, 0, 0), $X$(1/2, 0, 0), $T$(1/2, 1/2, 0), and $Y$(0, 1/2, 0). In the inset of (c), an energy surface is shown to better illustrate the band structure near the $\Gamma$ point.}
  \label{fig:band}
\end{figure}

Having obtained the optimized crystal structures and examined their energetics, we set to compute the associated band structures to determine which of the 14 carbon nanosheets are metals or semiconductors. Figure~\ref{fig:band} displays the calculated band structures with the PBE functional. We find that five types (types 1, 3, 8, 11, and 12) of the carbon nanosheets are semiconducting, one carbon nanosheet (type 9; graphene) is semimetallic with a Dirac cone, and the remaining eight nanosheets are metallic with valence bands crossing the Fermi level. The bandgaps of the semiconducting carbon nanosheets are listed in Table~\ref{summary}. We see that the bandgap sizes of the semiconducting carbon nanosheets span from 0.14 to 0.89 eV, which lie outside the visible light spectrum. But it is expected that these bandgaps are underestimated due to the usage of the PBE functional.\cite{PhysRevLett.51.1884} Advanced theoretical methods such as the hybrid density functional\cite{paier2006screened} and many-body theory\cite{PhysRev.139.A796} may correct the bandgaps to larger values that are useful for solar energy-related applications. 

The band structures of several of the 14 carbon nanosheets are worth of further description. The two valence bands with the highest energies are nearly degenerate throughout the entire specified $k$-point path. The bottom two conduction bands exhibit similar degeneracy. Because types 2 and 4 carbon nanosheets are equivalent, their band structures are identical. The conduction band minimum (CBM) and valence band maximum (VBM) of type 3 carbon nanosheet at the $\Gamma$ point exhibits a cone shape at first glance. To examine whether the conduction and valence bands form a Dirac cone as found in graphene,\cite{novoselov2005two} we calculate the energy surface using a uniform 81~$\times~$81 $k$-point grid near the $\Gamma$ point. The inset of Fig.~\ref{fig:band}(c) shows that type 3 carbon nanosheet displays a direct bandgap instead of a Dirac cone at the $\Gamma$ point. We also calculate the electron and hole effective masses to be 0.07 $m_0$ ($m_0$: electron rest mass). Such a small effective mass is beneficial for electrical transport. In contrast to the direct bandgap of type 3 carbon nanosheet, the bandgaps of types 1, 8, 11, and 12 are all indirect. For type 5 carbon nanosheet, we observe that several flat bands distributed around the Fermi level. These bands originate from the energy levels of the isolated C atoms in the centers of the hexagons. Similarly, most of the bands of type 11 carbon nanosheet show nearly flat dispersions, consistent with the structure consisting of isolated clusters. We additionally calculate the Fermi velocity $v_\mathrm{F}$ for the Dirac cone found in type 9 carbon nanosheet. We determine $v_\mathrm{F}$ as 8.2 $\times~10^5$ m/s, which is close to reported $\times~10^6$ m/s in the literature.\cite{PhysRev.71.622,RevModPhys.81.109} For type 11 carbon nanosheet, because the band gap is formed between nearly flat valence and conduction bands, the joint density of states is high that results in strong optical absorption.     

\section{Conclusions}
 In conclusion: We have carried out DFT calculations on carbon nanosheets using 14 distinct types of pentagons as the starting structures. We found that only one carbon nanosheet with the C atoms initially located at the vertices of a pattern of type 2 or 4 pentagons is capable of retaining the pentagonal pattern. We also showed attractive electrical properties such as small carrier effective masses that are attached to the predicted carbon nanosheets. We expect that the structural properties such as elastic modulus and cohesive energy to be calculated in the future work are useful to be included in the fitting processes of C-C empirical potentials in order to improve their transferrability. Another example of future work could be replacing carbon by other elements to explore the resulting 2D structures. Our theoretical work indicates that pentagons as well as other polygonal geometries offer a copious source of initial structures for DFT calculations, leading to an alternative method for discovering 2D materials that exhibit exotic properties for nanodevice applications.
\begin{acknowledgments}
We thank the Fulton Undergraduate Research Initiative (FURI) and the start-up funds from Arizona State University. This research used computational resources of the Texas Advanced Computing Center under Contracts No.TG-DMR170070. 
\end{acknowledgments}
\bibliography{references}
\end{document}